# Validation and Uncertainty Quantification for Wall Boiling Closure Relations in Multiphase-CFD Solver

Yang Liu[a,*] and Nam Dinh[b]

[a] North Carolina State University, Department of Nuclear Engineering
911 Oval DR, RM 2301, Engineering Building III
Raleigh, NC 27606

[b] North Carolina State University, Department of Nuclear Engineering
3145 Burlington Laboratories
Raleigh, NC 27695

[*]Email: yliu73@ncsu.edu

Number of pages: 37
Number of tables: 4
Number of figures: 10

**Abstract**

The two-fluid model based Multiphase Computational Fluid Dynamics (MCFD) has been considered as one of the most promising tools to investigate two-phase flow and boiling system for engineering purposes. The MCFD solver requires closure relations to make the conservation equations solvable. The wall boiling closure relations, for example, provide predictions on wall superheat and heat partitioning. The accuracy of these closure relations significantly influences the predictive capability of the solver. In this paper, a study of validation and uncertainty quantification (VUQ) for the wall boiling closure relations in MCFD solver is performed. The work has three purposes: i). identify influential parameters to the quantities of interest of the boiling system through sensitivity analysis; ii). evaluate the parameter uncertainty through Bayesian inference with the support of multiple datasets; iii). quantitatively measure the agreement between solver predictions and datasets. The widely used Kurul-Podowski wall boiling closure relation is studied in this paper. Several statistical methods are used, including Morris screening method for global sensitivity analysis, Markov Chain Monte Carlo (MCMC) for inverse Bayesian inference, and confidence interval as the validation metric. The VUQ results indicated that the current empirical correlations-based wall boiling closure relations achieved satisfactory agreement on wall superheat predictions. However, the closure relations also demonstrate intrinsic inconsistency and fail to give consistently accurate predictions for all quantities of interest over the well-developed nucleate boiling regime.

## I. Introduction

Two-phase flow and boiling heat transfer occurs in many situations and are used for thermal management in various engineered systems with high energy density, from power electronics to heat exchangers in power plants and nuclear reactors. Essentially, two-phase and boiling heat transfer is a complex multiphysics process, which involves different interactions between heated solid surface, liquid, and vapor, including nucleation, evaporation, condensation, interfacial mass/heat/momentum exchange. In current practices, the Eulerian-Eulerian two-fluid model based Multiphase Computational Fluid Dynamics (MCFD) has been considered as state-of-the-art method to model the two-phase flow with boiling in complex industrial applications with local details of flow and boiling features. Promising results from MCFD solver have been demonstrated in various applications [1-4].

The basis of the two-fluid model is the averaged field conservation equations, thus eliminating the need for tracking local instantaneous variables including time- and space-resolved interfaces to achieve computational efficiency. The averaging process, however, strips off significant information of the interfacial dynamics and details of inter-phase mass, momentum, and energy exchange processes. Thus, closure relations characterizing those phenomena need to be included to render the conservation equations solvable. Due to the complex nature of two-phase flow and boiling, those closure relations are still highly dependent on empirical parameters. The uncertainties introduced by the closure relations are still not well quantified, thus the predictive capability of the MCFD solver is hampered. Based on this fact, the Validation and Uncertainty Quantification (VUQ) of those closure relations are essential for improving the predictive capability of the MCFD solver.

Validation is important for model development and evaluation. Although validation is conducted in most of the model development works, it has two issues in many current practices. In many cases, a model is declared "validated" by comparing the quantities of interest (QoIs) predicted by solver and the experimental data on a graph. Such comparison, while provides a basic understanding of the model accuracy, cannot generate a comprehensive measure of the model-data discrepancy. Moreover, the model uncertainty and the experiment uncertainty are usually neglected in this type of treatment. The other issue comes from the usage of data, in many cases, the closure relations are "validated" based on different separate effect test (SET) data respectively. Those "validated" closure relations are then integrated into the solver to generate solutions and compare with the integral effect test (IET) data. Such "divide-and-conquer" approach is convenient and useful for a solver whose system can be decomposed into several components that have minor interactions. For highly non-linear solver (e.g. MCFD solver) whose components are highly coupled, such treatment could introduce significant uncertainty, since it neglects the interactions between the closure relations and the phenomena represented by them. One issue arising from such "divide-and-conquer" approach is inconsistency within the closure relations which is analyzed in this paper. When inconsistency exists in the model, the model fails to predict all the QoIs of it within acceptable uncertainty.

To address those two issues, the validation metrics [5-7] and the total-data-model-integration (TDMI) [8, 9] have been proposed by different researchers. The validation metric provides a quantitative and objective measurement of the agreement between solver predictions and datasets. The TDMI approach, on the other hands, aims at maximizing the information from multiple datasets to provide a better use of various (heterogeneous) data in the validation process of a complex solver. The general idea of TDMI is to treat the solver, closure relations and all available datasets in an integrated manner. In this paper, the Bayesian framework is applied to incorporate TDMI in the VUQ of MCFD solver. The Bayesian framework combines the prior subjective knowledge (in the form of parameters' prior distribution) and objective measurement (in the form of likelihood function) to generate a comprehensive uncertainty quantification on the key parameters of a solver. The difference between the traditional practices and the Bayesian based TDMI approach can be illustrated in Figure I.

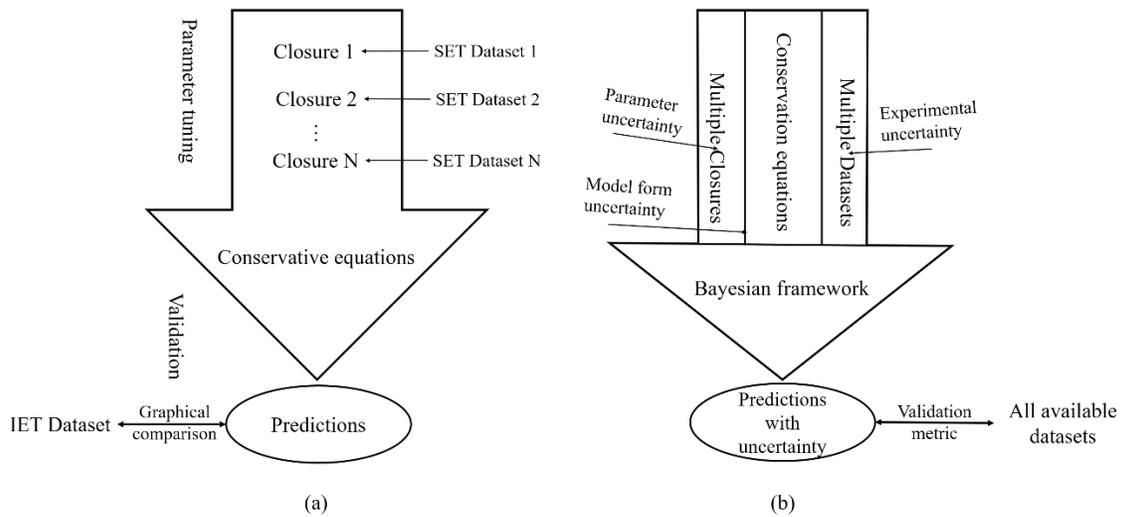

**Figure I. Two different VUQ practices: (a) Traditional practice; (b)VUQ based on TDMI and validation metric**

This paper formulates, implements and demonstrates a VUQ procedure which aims at providing a quantitative assessment of parameter uncertainties and the model-data discrepancy through the TDMI approach. In the work, the wall boiling closure relations of the MCFD solver are chosen as a demonstrative example of the VUQ procedure. The ultimate purpose of the VUQ procedure is to provide insights for a comprehensive evaluation of the predictive capability of a highly non-linear solver which includes multi-physics and/or multi-scale phenomena.

## II. Model description

The two-fluid model based MCFD solver relies on solving three ensemble averaged conservation equations for mass, momentum and energy. The k-phasic mass conservation equation can be written as

$$\frac{\partial(\alpha_k \rho_k)}{\partial t} + \nabla \cdot (\alpha_k \rho_k \mathbf{U}_k) = \Gamma_{ki} - \Gamma_{ik}, \tag{1}$$

where the two terms on the left-hand side represent the rate of change and convection, the two terms on the right-hand side represent the rate of mass exchanges between phases due to condensation and evaporation.

The k-phasic momentum equation is given by

$$\frac{\partial(\alpha_k \rho_k \mathbf{U}_k)}{\partial t} + \nabla \cdot (\alpha_k \rho_k \mathbf{U}_k \mathbf{U}_k) = -\alpha_k \nabla p_k + \nabla \cdot [\alpha_k(\tau_k + \tau_k^t)] + \\ \alpha_k \rho_k \mathbf{g} + \Gamma_{ki} \mathbf{U}_i - \Gamma_{ki} \mathbf{U}_k + \mathbf{M}_{ki}, \tag{2}$$

where $i$ represents the interphase between two phases.

The k-phasic energy conservation equation in terms of specific enthalpy can be given as

$$\frac{\partial(\alpha_k \rho_k h_k)}{\partial t} + \nabla \cdot (\alpha_k \rho_k h_k \mathbf{U}_k) = \nabla \cdot \left[\alpha_k \left(\lambda_k \nabla T_k - \frac{\mu_k}{\Pr_k^t} \nabla h_k\right)\right] + \\ \alpha_k \frac{Dp}{Dt} + \Gamma_{ki} h_i - \Gamma_{ik} h_k + q_k, \tag{3}$$

where the terms on the right-hand side represent heat transfer in phase $k$, work done by pressure, enthalpy change due to evaporation and condensation, and heat flux from the wall. The wall boiling heat transfer is modeled by a set of closure relations.

For a typical two-phase flow and boiling system, the nucleate boiling on the heated surface is one key process. The boiling involves complex multi-physics processes including interaction between vapor, liquid, and the wall surface, including nucleation, bubble growth, and bubble departure. Such process cannot be directly resolved in the MCFD solver, thus closure relations are introduced to describe the averaged effect of the process. The fundamental idea of the wall boiling closure relations that widely used in current boiling modeling practices was proposed by Kurul and Podowski [10] which relies on decomposing the applied wall heat flux into several components. Each component is modeled with one or several empirical correlations. The wall boiling closure relation combines all those empirical correlations to form a non-linear equation and solve it to obtain the QoIs related to the boiling process.

One of most widely used heat partitioning algorithms is to decompose the applied heat flux into three components [10]:

$$q_{Wall} = q_{Ev} + q_{Qu} + q_{Fc}. \tag{4}$$

The quenching heat transfer component $q_{qu}$ represents the heat transfer process occurs when bubble departures and cool liquid quenches the hot heating surface. Semi-analytical correlation with several empirical parameters has been developed for this term [11]:

$$q_{qu} = A_b \frac{2}{\sqrt{\pi}} f_d \sqrt{t_{wait} \lambda_l \rho_l c_{p,l}} (T_{sup} - T_l), \tag{5}$$

where $A_b$ in the expression is the effective bubble area fraction,

$$A_b = \max\left(\pi \left(a \frac{D_d}{2}\right)^2 N_a, 1.0\right), \tag{6}$$

and $a$ is the bubble influence factor which is an empirical parameter., $t_{wait}$ is the waiting time between the bubble departure and the appearance of a new bubble at a given nucleation site. In this work, the model proposed by [12] is selected:

$$t_{wait} = \frac{e}{f_d}, \qquad (7)$$

where $e$ is the waiting time coefficient, the original suggested value is 0.8.

The forced convective heat transfer component to the liquid phase can be expressed as:

$$q_{Fc} = (1 - A_b)h_l(T_{sup} - T_l), \qquad (8)$$

where $h_l$ is the convective heat transfer component which can be modeled by the near wall function of turbulence modeling. When assuming the flow in the near wall cell is within the logarithmic layer region, the corresponding liquid convective heat transfer coefficient can be modeled as:

$$h_l = u_\tau \rho c_p \left[ Pr_t \frac{1}{\kappa} \ln(Ey^+) + P \right]^{-1}, \qquad (9)$$

where $E$ are an empirical parameters from the wall function. $P$ is a function which governs the velocity at which the logarithmic and viscous regions of the thermal profiles intersect [13]:

$$P = 9.24 \left[ \left( \frac{Pr}{Pr_t} \right)^{3/4} - 1 \right] \left[ 1 + 0.28 \exp\left( \frac{-0.007 Pr}{Pr_t} \right) \right] \qquad (10)$$

The evaporation heat transfer component $q_{Ev}$ is expressed by the correlations to describe nucleation process, including the nucleation site density $N_a$, the bubble departure diameter $D_d$, and the bubble departure frequency $f_d$:

$$q_{Ev} = \frac{\pi}{6} D_d^3 \rho_v f_d N_a h_{fg}. \qquad (11)$$

Due to the complexity of the nucleation process, these correlations are heavily dependent on the empirical studies, various empirical correlations have been proposed by researchers. A few representative examples are summarized in Table I-III , while a more comprehensive review can be found in [14].

**Table I. Selected models for active nucleation site density**

| Model | Empirical correlation for $N_a$, $m^{-2}$ |
|---|---|
| Lemmert and Chawla [15] | $N_a = (aT_{\text{sup}})^b$, $a = 210$, $b = 1.805$ |
| Wang and Dhir [16] | $N_a = 5 \times 10^{-31}(1 - cos\theta)R_c^{-6.0}$ |
| Yang and Kim [17] | $N_a = N_{avg}\phi(\beta)exp(-CR_c)$ |
| Hibiki and Ishii [18] | $N_a = N_{avg}\left[1 - exp\left(-\dfrac{\theta^2}{8\mu_{con}^2}\right)\right]\left[exp\left(\dfrac{\lambda' g(\rho^+)}{R_c}\right) - 1\right]$ |

**Table II. Selected models for bubble departure diameter**

| Model | Empirical correlation for $D_d$, m |
|---|---|
| Cole and Rohsenow [19] | $D_d = 1.5 \times 10^{-4}\sqrt{\dfrac{\sigma}{g\Delta\rho}}\left(\dfrac{\rho_l C_{pl} T_{sat}}{\rho_g h_{fg}}\right)^{5/4}$ |
| Tolubinsky and Konstanchuk [20] | $D_d = \min[0.06 exp(-\Delta T_{sub}/45), 0.14]$, mm |
| Kocamustafaogullari [21] | $D_d = 1.27 \times 10^{-3}\left(\dfrac{\rho_l - \rho_g}{\rho_g}\right)^{0.9} d_{ref}$ |
| Zeng et al.[22] | Mechanistic model bubble departure/lift-off based on force balance analysis |

**Table III. Selected models for bubble departure frequency**

| Model | Empirical correlation for $f$, $s^{-1}$ |
|---|---|
| Cole[23] | $f_d = \sqrt{\dfrac{4g(\rho_l - \rho_g)}{3D_d \rho_l}}$ |
| Kocamustafaogullari and Ishii [24] | $f_d = \dfrac{1.18}{D_d}\left[\dfrac{\sigma g(\rho_l - \rho_g)}{\rho_l^2}\right]^{0.25}$ |
| Podowski et al. [12] | Mechanistic model accounts for waiting time and bubble growth time |

## III. VUQ workflow

In this paper, a general workflow for VUQ on wall boiling model is proposed which couples state-of-art UQ methods and MCFD solver, with the support of high fidelity multi-scale database. This VUQ workflow can serve for three purposes: i). identify influential parameters/phenomena for a given QoI; ii). evaluate the parameter uncertainty and the QoI uncertainty; iii). identify the potential inconsistency within the boiling closure relations. Those outputs can be generalized and reviewed to give a comprehensive evaluation of the predictive capability of the solver.

In this paper, the MCFD solver developed based on open source CFD library OpenFOAM is used [25] . On the other hand, the proposed VUQ procedure is based on non-intrusive UQ methods, thus no modification of the simulation code is required for its application. This makes it applicable to commercial MCFD solver such as STAR-CCM+. However, it needs to point out that the proposed VUQ framework is based on the TDMI idea, thus a corresponding support of high fidelity multi-scale databases for the QoIs is required. In this work, the wall boiling model in MCFD solver is chosen as an example of the framework, the high-resolution boiling data serves as database support. The scheme of the VUQ workflow is shown in Figure II. The components of the framework are introduced in more detail in this section.

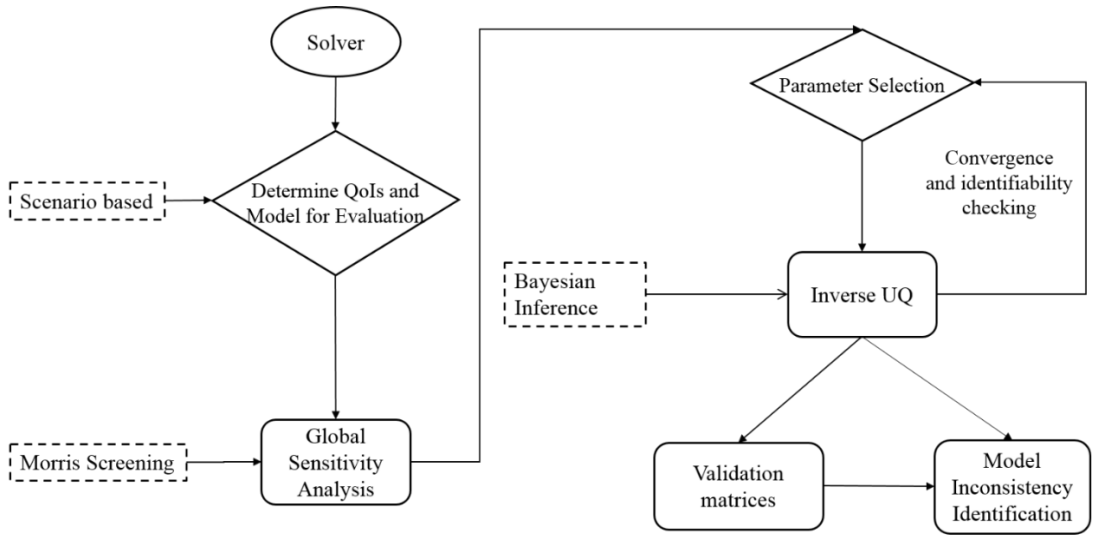

**Figure II. Workflow of the proposed VUQ process**

**II.A Determine QoIs and model for evaluation**

The first step is to determine the QoIs and the corresponding model for VUQ evaluation based on the scenario interested. Taking the subcooled flow boiling problem discussed in this paper as an example, the major QoIs are the heat transfer in the heated wall as well as the wall superheat. Thus, the wall boiling model would be the evaluation model subject to VUQ. Datasets for all (or at least a large fraction of) the QoIs with known uncertainty are desired for the VUQ application. The evaluation model needs to be analyzed to identify its structure, its connection with empirical closures, and the empirical parameters employed in the closures. The prior uncertainty distribution of those parameters needs to be defined. In this paper, the non-informative priors to these parameters are chosen, the range is dependent on expert judgment as well as from previous simulation practices from different researchers.

**II.B Sensitivity analysis**

The purpose of conducting Sensitivity Analysis (SA) in this paper is to identify the influential parameter and specify regimes in the parameter space that has the highest impact on the QoIs. There are two types of SA: the local SA and global SA. In this paper, the global SA is performed and the Morris Screening method is used [26]. Morris screening method evaluates the sensitivity of parameters over the whole input space through a one-step-at-a-time approach, which means in each model evaluation, only one parameter is given a new value. Morris Screening method can provide the capacity to rank parameters according to their importance but cannot quantify the contribution of a parameter to the output variability of a QoI. Compared to other global SA method such as variance-based decomposition, the major advantage of Morris Screening is its low computational cost. Morris Screening bases on the linearization of the model. To construct the elementary effect, one partitions [0,1] into $l$ levels. Thus the elementary effect associated with the $i^{th}$ input can be calculated as

$$d_i(\boldsymbol{\theta}) = \frac{y^M(\theta_1, \dots, \theta_{i-1}, \theta_i + \Delta, \theta_{i+1}, \dots \theta_p) - y^M(\boldsymbol{\theta})}{\Delta}, \tag{12}$$

where the step size $\Delta$ is chosen from the set

$$\Delta \in \left\{\frac{1}{l-1}, \dots, 1 - \frac{1}{l-1}\right\}. \tag{13}$$

For $r$ sample points, the sensitivity measurement for $x_i$ can be represented by the sampling mean $\mu_i$, standard deviation $\sigma_i^2$, and mean of absolute values $\mu_i^*$, which can be calculated respectively.

$$\mu_i^* = \frac{1}{r} \sum_{j=1}^{r} |d_i^j(\mathbf{q})| \tag{14}$$

$$\mu_i = \frac{1}{r} \sum_{j=1}^{r} d_i^j(\mathbf{q}) \tag{15}$$

$$\sigma_i^2 = \frac{1}{r-1}\sum_{j=1}^{r}(d_i^j(\mathbf{q}) - \mu_i)^2 \qquad (16)$$

The mean represents the effect of the specific parameter on the output, which the variance represents the combined effects of the input parameters due to nonlinearities or interactions with other inputs. The obtained μ, μ* and σ can help ranking the parameters by relative order of importance. If the value of sigma is high compared to mu (same order of magnitude), a non-linear influence and/or interactions with other parameters are detected. This measure is however only qualitative.

It should be noted that while the global SA reveals the general trend of parameter's importance over the whole parameter space, the detailed information about the parameter sensitivity at a given value can only be obtained from local SA. One example for local SA is the adjoint method, which has wide applications in reactor neutronics [27, 28]. It should also be noted that recent advancements have brought this method in the area of reactor thermal-hydraulics [29].

**II.C Parameter selection**

For a complex system such as the wall boiling closures in MCFD solver, the parameter identifiability arises as a major issue for conducting Bayesian inference. This issue stems from the fact that with a limited number of datasets, there could exist different values of parameters that produce very similar results and fit the data equally well. The convergence of Bayesian inference would face difficulty with non-identifiable parameter exists. On the other hand, it is desired to find a subset of parameters that not only can be identified against each other but also with high sensitivity so that the Bayesian inference would give smaller posterior uncertainties of parameters. In this sense, the parameter selection is tightly correlated with SA. The objective of this step is to perform SA and select a subset of parameters based on SA for the Bayesian inference in next step.

It is natural to select parameters based on the global SA results. The general idea is to select parameters with high impact to the QoIs while do not have identifiability issue. However, even two highly influential parameters cannot guarantee that they can be identified with each other. Thus based on the global SA results, the parameter selection is an *ad hoc* solution which requires several times of trial-and-error. The basic idea is to include all the influential parameters to do a preliminary try on Bayesian inference, if identifiability issue exists, get rid of one parameter with least importance and do another trial. A subset of influential and identifiable parameters can be selected with the test of all possible combinations of influential parameters.

**II.D Uncertainty quantification with Bayesian inference**

The uncertainty quantification step uses Bayesian inference to inversely quantify the parameter uncertainty with the available datasets, then propagate the parameter uncertainty through the solver to evaluate the uncertainty of QoIs predicted by the solver.

For any computational model, the relationship of its outputs to experimental datasets can always been expressed as:

$$\mathbf{y}^E(\mathbf{x}) = \mathbf{y}^M(\mathbf{x}, \boldsymbol{\theta}) + \delta(\mathbf{x}) + \varepsilon(\mathbf{x}), \quad (17)$$

where $\mathbf{y}^E(\mathbf{x})$ is the experimental measurement, $\mathbf{y}^M(\mathbf{x}, \boldsymbol{\theta})$ is the model prediction, $\delta(\mathbf{x})$ is the model bias term, which is usually caused by missing physics, simplified assumptions or numerical approximations in the model, and $\varepsilon(\mathbf{x})$ is the measurement uncertainty. In this paper, $\varepsilon(\mathbf{x})$ is assumed to be i.i.d (independent and identically distributed) normal distributions with zero means and know variance $\sigma^2$:

$$\varepsilon(\mathbf{x}) \sim N(0, \sigma^2 \mathbf{I}) . \tag{18}$$

It worth mention that although zero mean normal distribution of measurement uncertainty is reasonable and has been widely used for most cases, such an assumption may become questionable for some problems. A more sophisticated treatment is termed "full Bayesian" which assumes the measurement uncertainty follows a certain distribution which is controlled by hyperparameters. These hyperparameters are assigned with prior distributions and are considered in the Bayesian framework along with other physical parameters. A more detailed explanation of this approach can be found in [30].

An assumption is made in this work which assumes the model bias $\delta(\mathbf{x})$ is small enough compared to the parameter uncertainty and thus can be negligible in the following work. The inverse UQ process aims to evaluate the uncertainty of the parameter based on the data. In the framework of Bayesian inference, which treats the parameter as random variables, the prior knowledge for the parameter is also considered. The prior knowledge usually comes from previous simulations, other experimental observations or purely expert opinion. The Bayes formula is the foundation of Bayesian inference:

$$p(\boldsymbol{\theta}|\mathbf{y}^E) = \frac{p(\mathbf{y}^E|\boldsymbol{\theta})p_0(\boldsymbol{\theta})}{p(\mathbf{y}^E)} \propto p(\mathbf{y}^E|\boldsymbol{\theta})p_0(\boldsymbol{\theta}), \tag{19}$$

where $p(\boldsymbol{\theta}|\mathbf{y}^E)$ is the posterior distribution of the empirical parameter. $p(\mathbf{y}^E|\boldsymbol{\theta})$ is the conditional probability that given the values of empirical parameters, the probability of obtaining the specific experimental measurement. In Bayesian inference, the conditional probability is converted to the likelihood function $L(\boldsymbol{\theta}|\mathbf{y}^E)$, which can be expressed as

$$L(\boldsymbol{\theta}|\mathbf{y}^E) = \frac{1}{(2\pi\sigma^2)^{n/2}} \exp\left(-\sum_{i=1}^{n} \frac{\left(y^E(x_i) - y^M(x_i, \boldsymbol{\theta}) - \delta(x_i)\right)^2}{2\sigma^2}\right). \tag{20}$$

Directly evaluating the posterior distribution through the Bayes formula is extremely challenging, especially for high dimensional parameters $\boldsymbol{\theta}$. An widely used alternative to construct the $p(\boldsymbol{\theta}|\mathbf{y}^E)$ is the method of Markov Chain Monte Carlo (MCMC) [31], which constructs Markov chains whose stationary distribution is the posterior distribution $p(\boldsymbol{\theta}|\mathbf{y}^E)$.

Once $p(\boldsymbol{\theta}|\mathbf{y}^E)$ is obtained, the distribution can be propagated through the solver, the uncertainty distributions of the QoIs predicted by the solver are then obtained.

**II.E Confidence interval evaluation**

Validation should be performed with the QoIs obtained through the UQ process. In previous validation practices, a widely used approach is to graphically compare the model predictions with the experimental measurement. Such "graphical comparison", while provides a basic understanding of the model accuracy, cannot generate a quantitative measurement of the simulation-data agreement, and can hardly lead to a reasonable evaluation of the solver. To address this issue, the validation metrics that aim to provide an objective and quantitative measurement of the agreement between model predictions and experimental measurement are proposed. According to [32], the validation metrics can be characterized into three types: the hypothesis testing, the confidence interval, and the area metric based validation metrics such as p-box. Among these metrics, the confidence interval provides a clear physical interpretation of the metric and is easy to implement. Thus, in this paper, the confidence interval is used as the validation metric.

Suppose there are $n$ times of repeated measurements of a certain QoI $\mathbf{y}_i^E$ under a given condition, the corresponding solver prediction is $\mathbf{y}^M$, the confidence interval can be calculated as

$$\left(\tilde{\mathbf{E}} - t_{\alpha/2,v} \cdot \frac{s}{\sqrt{n}}, \tilde{\mathbf{E}} + t_{\alpha/2,v} \cdot \frac{s}{\sqrt{n}}\right), \tag{21}$$

where $\tilde{\mathbf{E}}$ is the discrepancy between solver prediction $\mathbf{y}^M$ and mean of experimental measurements

$$\tilde{\mathbf{E}} = \mathbf{y}^M(\mathbf{x}, \boldsymbol{\theta}) - \sum_{i=1}^{n} \mathbf{y}_i^E, \tag{22}$$

$s$ is the standard deviation of the experimental data, $t_{\alpha/2,v}$ is the 1-α/2 quantile of the t-distribution with freedom of $v$ used to quantify the uncertainty of experimental data. The obtained confidence interval can be interpreted as "we have $(1 - \alpha) \times 100\%$ confidence that the true discrepancy between model and observed data is within the interval". In most practices, the $\alpha$ is chosen to be 0.05.

In this paper, the experimental data is collected from published paper [33], thus the detailed information of the measurement uncertainty analysis, such as the times of repeated measurement, is not available. To overcome this, the normal distribution is used to represent the uncertainty of experimental measurement

$$\left(\tilde{\mathbf{E}} - z_{\alpha/2} \cdot \sigma, \tilde{\mathbf{E}} + z_{\alpha/2} \cdot \sigma\right), \tag{23}$$

where $z_{\alpha/2}$ is the 1-α/2 quantile of the normal distribution, $\sigma$ is the standard deviation of the measurement which is available in the publication.

One disadvantage of the confidence interval is that it does not consider the QoIs' full uncertainty. A more comprehensive formulation of validation metrics is to treat the experimental measurement and the simulation results as both probability distributions and calculate the area metric [6]. On the other hand, the confidence interval calculated in this paper does reveal a similar information as the area metric with much simpler calculation:

the distance between a point estimate (simulation QoIs results) and a probability distribution (experimental data).

**II.F Prediction consistency identification**

For modeling a complex system such as boiling, the "divide-and-conquer" is an efficient modeling strategy. On the other hand, this approach neglects the possible interaction between different phenomena of the system. Such negligence, along with the simplification assumption in the modeling process, could result in inconsistency of the model. A model with such inconsistency will not be able to give the accurate prediction on all QoIs over multiple conditions. In this work, the confidence interval is also used to identify such inconsistency. The general idea is to compare the difference between i). confidence intervals for all QoIs from UQ performed with one experimental condition and ii). confidence intervals for all QoIs from UQ performed with all available experimental conditions. A large discrepancy between the two predictions suggests there is inconsistency within the closure relations since the large discrepancy indicates that there does not exist a set of parameter values for the closure relations that can accurately predict all the experimental observations simultaneously. The procedure to identify the inconsistency can be described as follows a). Perform UQ through Bayesian inference for QoIs under each experimental measurement separately, propagate the obtained parameter uncertainty through solver to obtain the uncertainty of QoIs, denoted as $y_{single}^j$, where $j$ stands for the measurement under $j^{th}$ input condition.

b). Construct confidence intervals for QoIs $y_{single}^j$, the obtained confidence interval

under $j^{th}$ input condition is denoted as $CI_{single}^{j}$

c). Perform Bayesian inference for QoIs under all conditions simultaneously, propagate the obtained parameter uncertainty through solver to obtain the uncertainty of QoIs under each condition, denoted as $y_{full}^{j}$ d). Construct confidence intervals for QoIs $y_{full}^{j}$, the obtained confidence interval under $j^{th}$ input condition is denoted as $CI_{full}^{j}$ e). Comparing $CI_{full}^{j}$ and $CI_{single}^{j}$, if significant difference in $j^{th}$ input condition is observed between these two intervals, then inconsistency within the closure relations with regard to $j^{th}$ input condition is identified. The inconsistency can be quantitatively evaluated by the distance or overlapping area between the two intervals.

## IV. Problem setup

The scenario studied in this paper is the subcooled flow boiling in the vertical channel. The data used in this work are reported in [33], which use cutting-edge diagnostics to measure detailed boiling processes and post-processing the results to generate data that compatible to MCFD solver, including averaged wall superheat and heat portioning over the whole heating surface. In this work, the experimental data under mass flux 500 kg/m²s and various heat fluxes are extracted and used in the inverse UQ process. The heat fluxes studied in this paper ranging from 500 kW/m² to 2400 kW/m², which basically cover the well-developed nucleate boiling regime. The detailed description regarding the experimental condition can be found in the original publication. Six datasets of different heat fluxes, including the wall superheat and three heat partitioning components, are extracted and used in the VUQ process.

It should be noted that although there are multiple sets of closure relations in a MCFD solver, such as interfacial forces, turbulence model, etc., only the wall boiling closure relations are considered in this work. Such consideration is based on two reasons. The first is that considering more closure relations would introduce more parameters. A large number of parameters will suffer from the "curse of high dimensionality", which will cause convergence problem for the Bayesian inference. More importantly, it is observed that perturbing parameters of closure relations other than wall boiling only has a small influence on the boiling heat transfer predictions of MCFD solver. In this sense, it is reasonable to consider only parameters in wall boiling closure relations for the VUQ of boiling heat transfer predictions of MCFD solver.

The wall boiling closure relations studied in this work are discussed in Section II from Eq.(4) to Eq.(10). The nucleation site density correlation studied in this work is proposed by Hibiki and Ishii [18]

$$N_a = N_{avg}\left[1 - exp\left(-\frac{\theta^2}{8\mu_{con}^2}\right)\right]\left[exp\left(\frac{\lambda' g(\rho^+)}{R_c}\right) - 1\right], \tag{24}$$

where

$$R_c = \frac{2\sigma\{1 + (\rho_g/\rho_f)\}/P_f}{exp\{h_{fg}(T_g - T_{sat})/R\,T_g T_{sat}\} - 1}, \tag{25}$$

$$g(\rho^+) = -0.01064 + 0.48246\rho^+ - 0.22712\rho^{+2} + 0.05468\rho^{+3}, \tag{26}$$

$$\rho^+ = \log\left(\frac{\rho_l - \rho_g}{\rho_g}\right). \tag{27}$$

Here $N_{avg}$, $\mu_{con}$ and $\lambda'$ are empirical parameters that represent the average cavity density, angle scaler, and cavity radius scaler respectively.

The bubble departure diameter correlation studied in this work is proposed by

Kocamustafaogullari [21]:

$$D_d = d_1 \theta \left(\frac{\sigma}{g\Delta\rho}\right)^{0.5} \left(\frac{\Delta\rho}{\rho_g}\right)^{0.9}. \qquad (28)$$

The bubble frequency correlation studied in this work is proposed by Kocamustafaogullari and Ishii [23]

$$f_d = \frac{f_{con}}{D_d}\left[\frac{\sigma g(\rho_l - \rho_g)}{\rho_l^2}\right]^{0.25}, \qquad (29)$$

where $f_{con}$ is the departure frequency constant whose suggested value is 1.18 .

**Table IV. Prior uncertainties of empirical parameters**

| Empirical parameter | Physical meaning | Nominal value | Lower bound | Upper bound |
|---|---|---|---|---|
| $N_{avg}$ | Nucleation site density coefficient | $4.72 \times 10^5$ | $4.72 \times 10^4$ | $4.72 \times 10^6$ |
| $\mu_{con}$ | Contact angle scaler | 0.722 | 0.4 | 3.14 |
| $d_1$ | Departure diameter constant | 0.0015 | 0.0005 | 0.003 |
| $f_{con}$ | Departure frequency constant | 1.18 | 0.5 | 3 |
| $a$ | Effective bubble area factor | 1 | 0.5 | 2 |
| $e$ | Bubble growth waiting time factor | 0.8 | 0.65 | 0.95 |
| $E$ | Wall function Log law offset | 9.73 | 1 | 15 |

Before performing the SA and UQ, the prior uncertainties of empirical parameters should be determined. In this paper, these prior uncertainties and the nominal values are prescribed based on expert judgment, the results are summarized in

Table IV. The uniform distributions are set for all parameters as the "non-informative" prior.

## V. Results and discussions

### V.A Sensitivity analysis and parameter selection

In this work, the Morris Screening method is used for SA. Two selected results of different heat fluxes for all the four QoIs are depicted in Figure III and Figure IV.

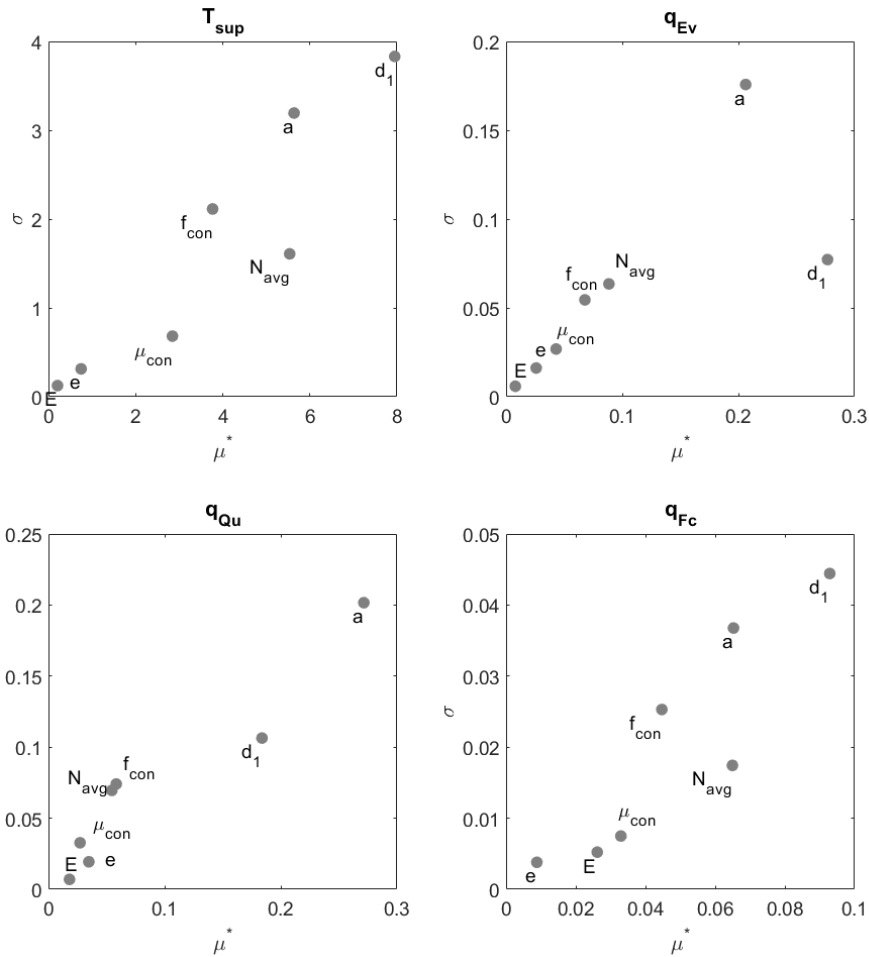

**Figure III. Morris screening measures for 4 QoIs ($q_{wall} = 500\ kW/m^2$)**

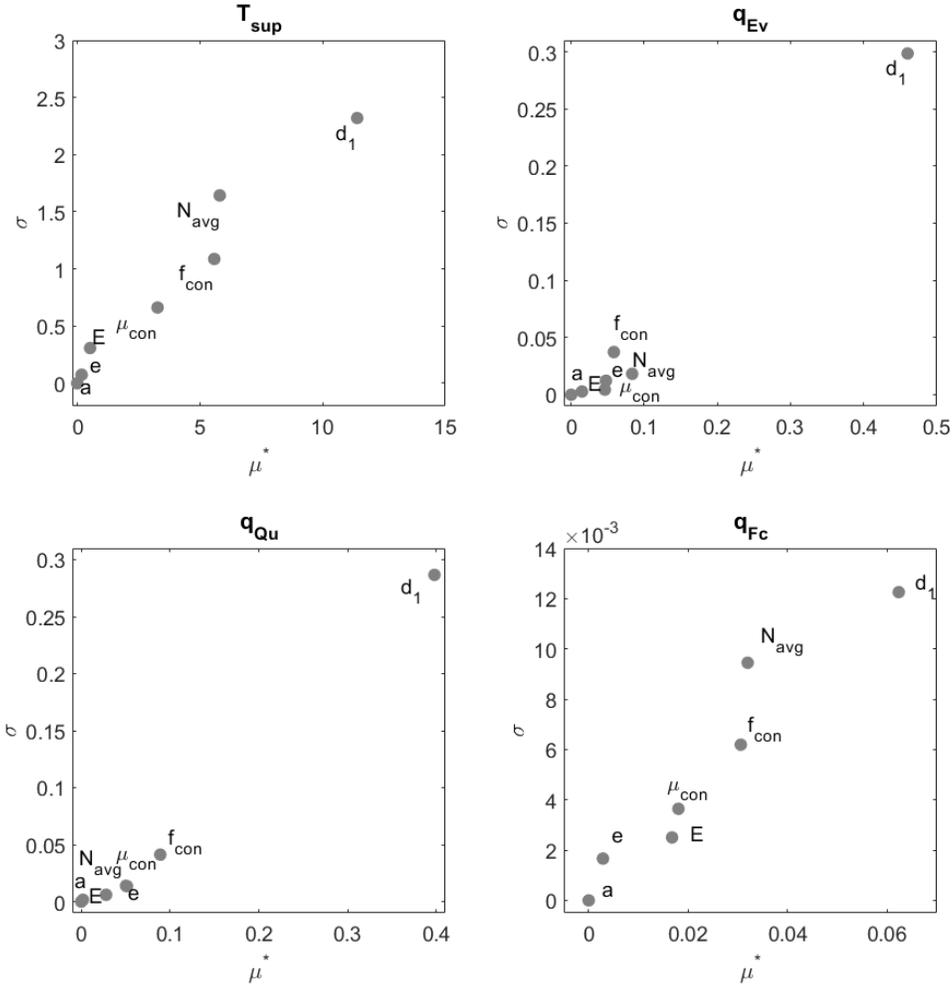

**Figure IV. Morris screening measures for 4 QoIs ($q_{wall} = 2400 \ kW/m^2$)**

In the plot, the x-axis represents the absolute value of the sampling mean $\mu_i^*$, the higher $\mu_i^*$ indicates the stronger influence of the parameter on the QoI. While the y-axis represents the standard deviation $\sigma_i$, higher $\sigma_i$ indicates stronger interaction of the parameter with other parameters. Thus, the parameters with both high $\mu_i^*$ and $\sigma_i$ (on the top right of the figure) are considered to be most influential to the QoIs.

It can be found from the two figures that the bubble departure diameter constant $d_1$ is an influential parameter for all QoIs in both heat fluxes. The effective bubble area factor $a$ is influential in the low heat flux case, but has trivial effect on the high heat flux case. This observation meets the expectation that in high heat flux the bubble area would influence nearly all the heating surface which means the effect area would be 1 no matter what value $a$ is. Generally speaking, the departure frequency constant $f_{con}$, the contact angle scaler $\mu_{con}$, the wall function coefficient $E$, and the nucleation site density coefficient $N_{avg}$ are influential to certain QoIs in both cases. To ensure consistency in following UQ and validation metric evaluation, a fixed combination of parameters are selected based on the averaged SA results. The averaged Morris screening measures over all the input heat fluxes are depicted in Figure V.

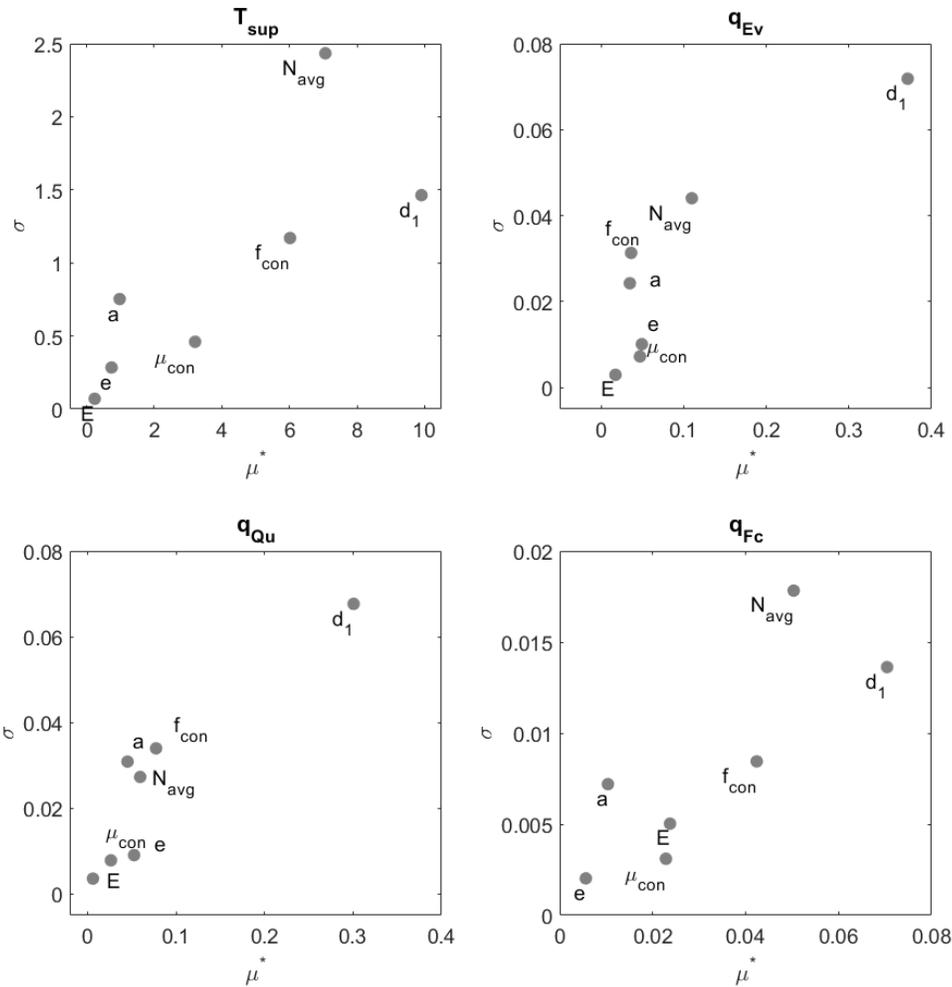

**Figure V. Morris screening measures for 4 QoIs (averaged on all heat fluxes)**

Through a few attempts of trial-and-error, a subset of influential and mutually identifiable parameters is selected, i.e. the effective bubble area factor $a$, the bubble departure diameter constant $d_1$, and the wall function log-law offset parameter $E$. These parameters are included in the following uncertainty analysis work.

**V.B Uncertainty quantification**

Directly inferring the posterior distribution of parameters through Bayes formula for a complex model with multiple parameters is extremely difficult. The MCMC method is an alternative for Bayesian inference and has demonstrated its applicability in thermal-hydraulics problems[34]. The general idea of MCMC is to construct Markov chains that converge to the posterior parameter distributions. For a given parameter, it is proved that the stationary distribution of the Markov chains is the posterior density. There are multiple algorithms for MCMC sampling; in this work, the Delayed Rejection Adaptive Metropolis (DRAM) algorithm [35] is chosen for MCMC. There are two features of DRAM, one is delayed rejection, which means if a candidate is rejected in the sampling process, an alternate candidate is constructed to induce greater mixing. The other is adaption, which means the covariance matrix of the parameters is continuously updated using the accepted candidates. MCMC will construct stationary distribution of a Markov chain that equals to the posterior distribution of the parameter.

The processed sample chains of all selected parameters and their autocorrelations are plotted in Figure VI. Good mixing and the fast decay of auto-correlations for all parameters can be observed which indicate the obtained samples can be regarded as the stationary distributions of the Markov chains.

The marginal and pair-wise joint distributions of the three parameters are plotted in Figure VII. The correlation between two coefficients is visible in the joint distribution. It can be found from the figure that there is very weak positive correlation between $a$ and $d_1$, but generally speaking, the three parameters can be regarded as independent with each other and thus are mutually identifiable.

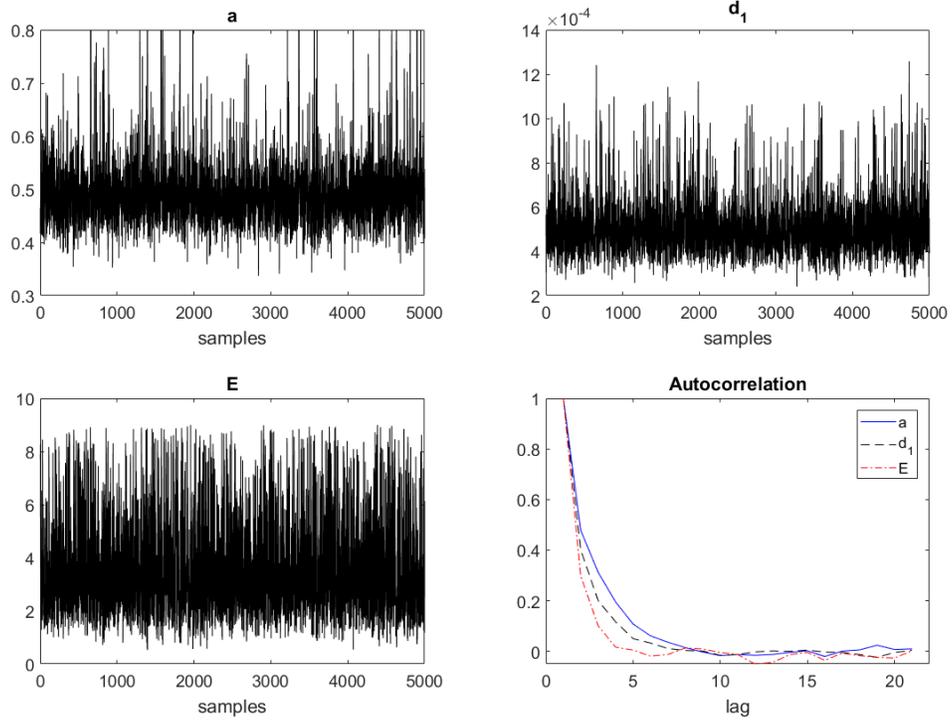

**Figure VI. MCMC sample traces and auto-correlations of studied parameters**

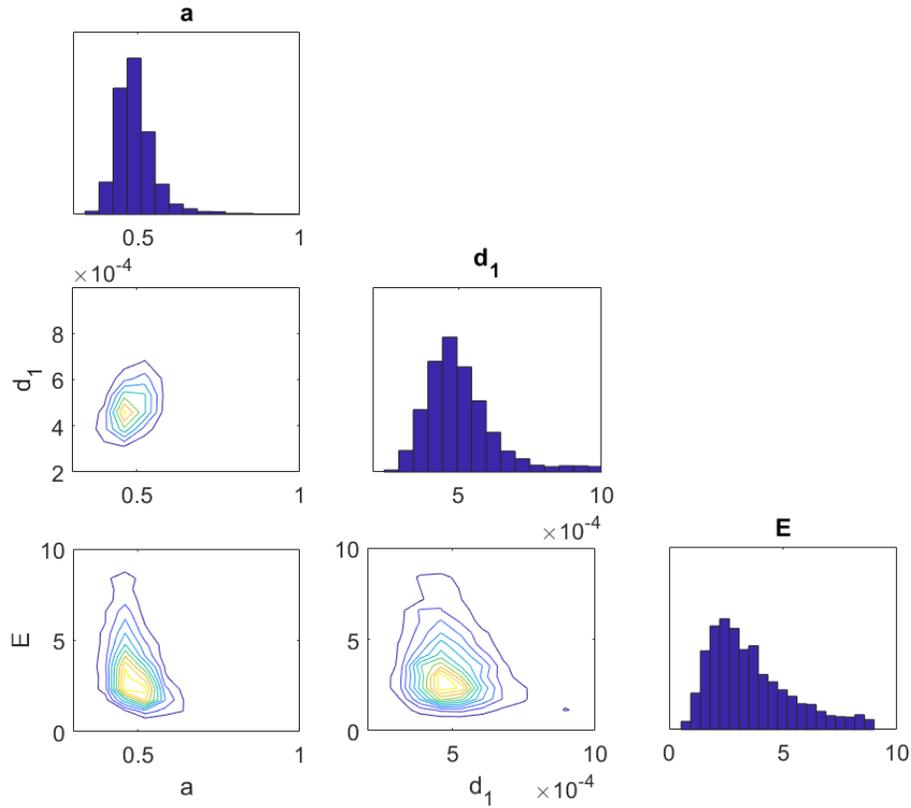

**Figure VII. Marginal and pair-wise joint posterior distributions of studied parameters**

For evaluating the prediction inconsistency, the MCMC has been performed 7 times to obtain 7 different sets of parameter posterior distributions: one case takes all the datasets to perform the MCMC, while each of the other 6 cases take only one dataset respectively. The results are depicted in Figure VIII.

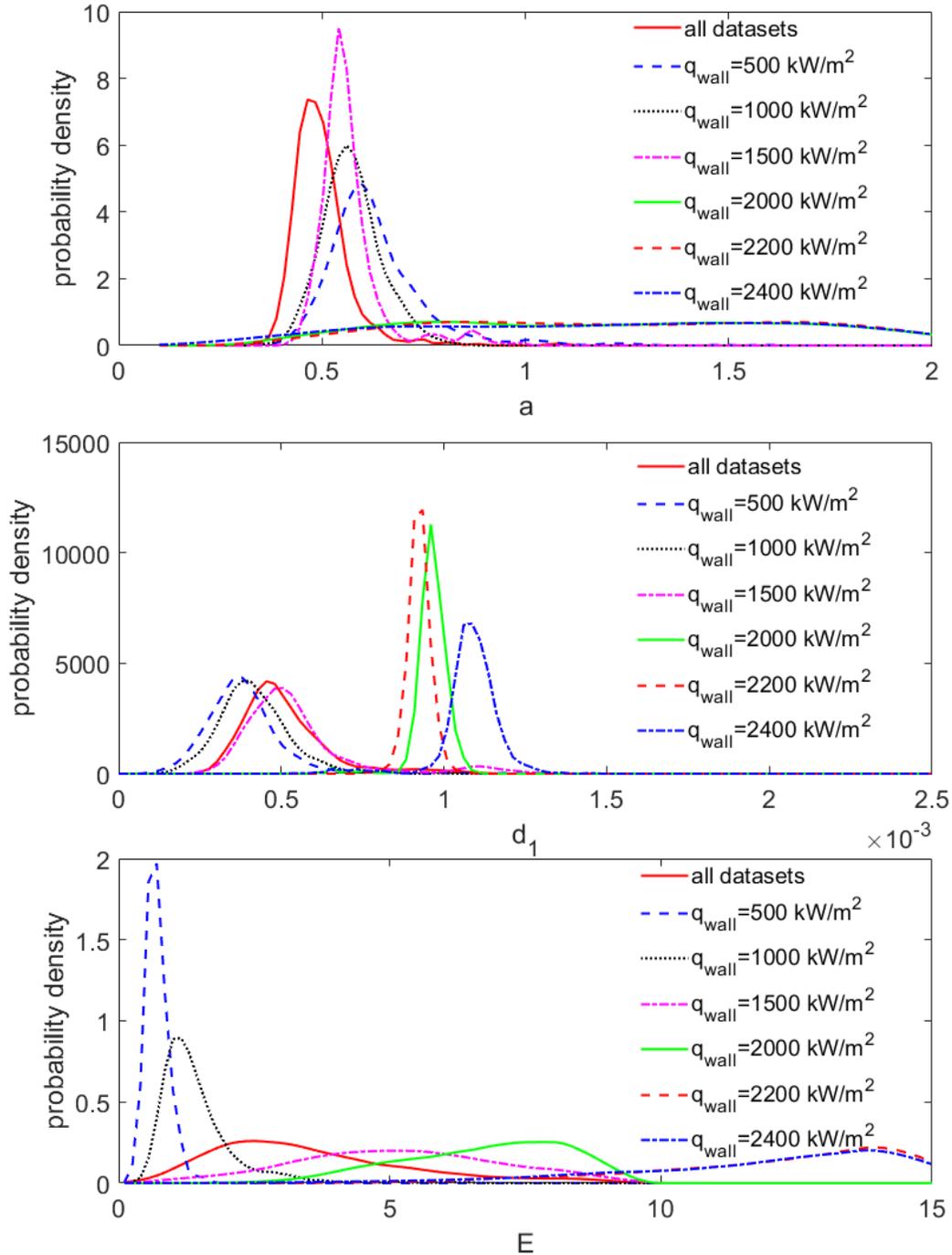

**Figure VIII. Marginal posterior distributions of parameters for different cases**

It can be found from Figure VIII that the $d_1$ is sensitive to the data in all cases and has been updated from flat uniform prior to sharply peaked distribution in all cases. $a$ is sensitive to relative low heat fluxes but are not sensitive in case where heat flux exceeds 2000 kW/m². $E$ is well informed by data to form sharply peaked distribution in low heat fluxes but are only weakly informed in heat flux exceeds 1000 kW/m$^2$. The reason is that $E$ only influences the forced convective heat transfer and this component contributes a less proportion of heat transfer in high flux regime. These observations are consistent with the Morris screening measures.

One issue that can be identified from the figure is that the distribution for different cases deviates significantly. This means the Bayesian inference cannot generate consistent posterior distributions with data of different conditions. Such deviations suggest there is inconsistency within the studied boiling closure relations. Due to this inconsistency, there does not exist a universal combination of parameters that can predict the whole well-developed nucleate boiling regime with good accuracy. In real practices, researchers are tuning the parameters to generate good agreement for their specific cases. Such tuning process, however, could be problematic when being extended to other conditions.

Another issue that should be noticed that $E$ is a parameter in the wall function of turbulence model which also has a significant influence on the near wall flow features. The widely accepted value for $E$ in turbulence modeling is around 9. This suggests the inconsistency could not only exist within the boiling closure relations but also exists between boiling closure relations and turbulence models. Considering the MCFD solver has tightly coupled closure relations of different categories, it is desired to treat all these relations simultaneously and evaluate their influence and interaction on different QoIs. However, the VUQ practice with all closure relations is still limited by the data availability at this time.

**V.C Inconsistency evaluation**

Once the posterior distributions of the parameters are obtained, they can be propagated through the solver to obtain the uncertainty of the QoIs. The six cases that take only one dataset each can combine together to generate the solver prediction $y_{single}$ ($y$ can be any QoI) as noted in Section II.F, while the case that takes all datasets into simultaneous consideration generates $y_{full}$. The comparison between experimental measurement and these two predictions are depicted in Figure IX.

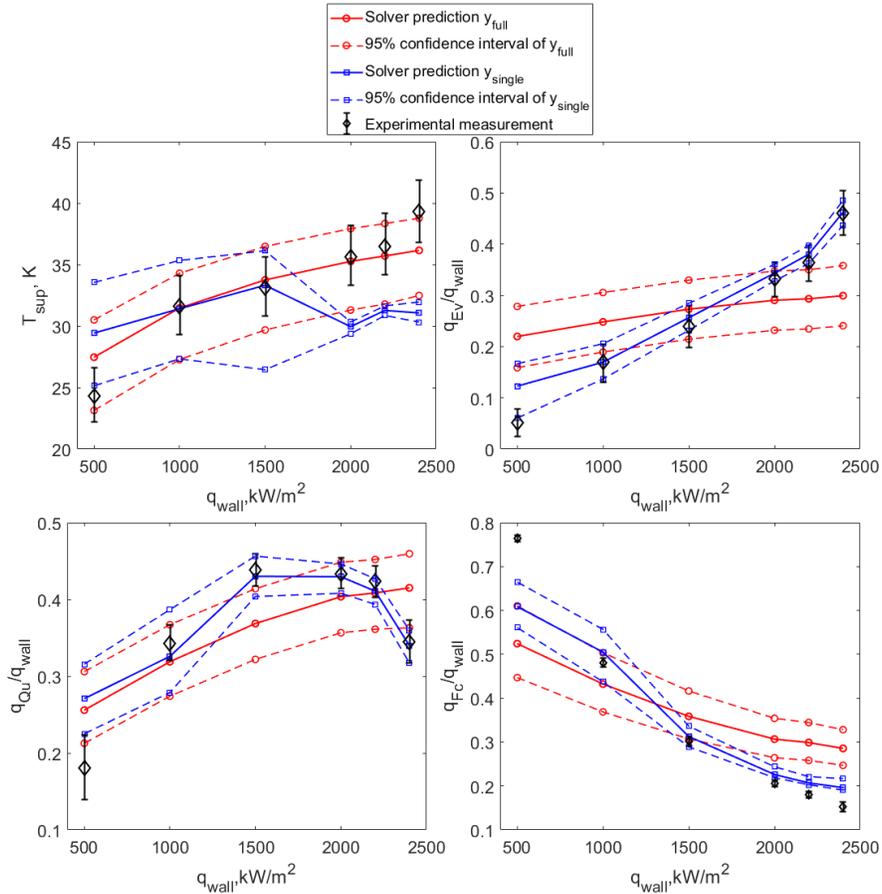

**Figure IX. Comparison of experimental measurements and solver predictions of two VUQ approaches**

It can be observed from Figure IX that the prediction of wall superheat $T_{sup}$ demonstrate satisfactory agreement with the experimental measurement. However, significant results discrepancy for other three QoIs between the two different VUQ approaches can be observed. The $y_{full}$ demonstrate monotonic behavior, thus it failed to capture the trend of $Q_{Qu}$ which first increases with heat flux then decreases in high heat flux regime. But the $\boldsymbol{y}_{single}$ captures such trends accurately. Moreover, except for the wall superheat predictions at high heat fluxes, the $\boldsymbol{y}_{single}$ demonstrated better agreement with the experimental measurement, compared to $\boldsymbol{y}_{full}$. This meets the experiences we gained in running MCFD simulations. In such simulations, the parameter can be carefully tuned to generate good agreement with experimental measurement. However, judging from the results shown in Figure IX, such tuned values may not be able to extend to other conditions.

The confidence intervals of the errors between solver predictions and datasets are depicted in Figure X. The interval closes to or covers zero indicate good agreement between solver predictions and experimental measurement. It can be found $\boldsymbol{y}_{full}$ has larger error interval for almost all cases compared to $\boldsymbol{y}_{single}$. Again, this confirmed that $\boldsymbol{y}_{single}$ has better agreement with the experimental measurement compared to $\boldsymbol{y}_{full}$.

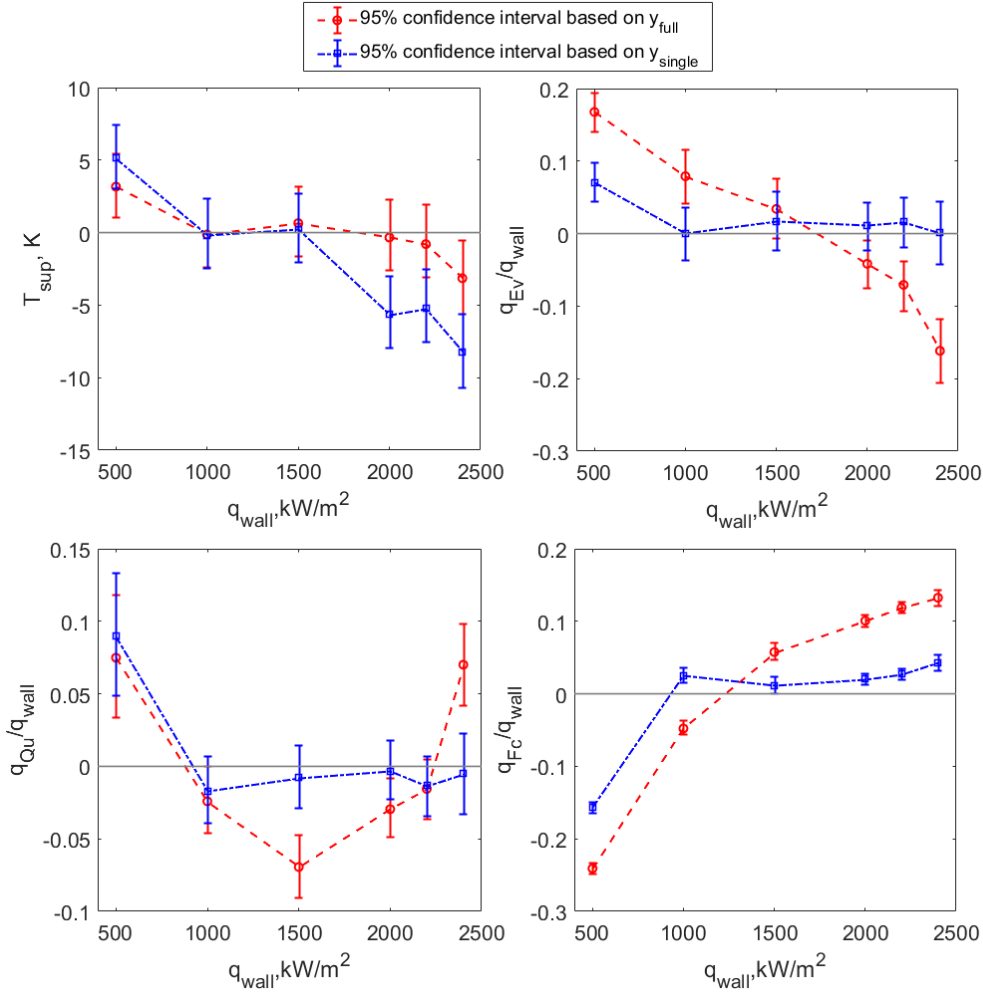

**Figure X. Prediction inconsistency identification using two different confidence intervals**

## VI. Conclusions

In this paper, validation and uncertainty quantification (VUQ) is performed for the wall boiling closure relations in MCFD solver. A VUQ procedure is proposed for the assessment of boiling closure relations which have multiple QoIs. Several statistical methods are used in this VUQ work, including global sensitivity analysis, Bayesian inference with Markov Chain Monte Carlo (MCMC), and confidence interval which serves as the validation metric to provide a quantitative and objective measurement for the agreement between solver predictions and datasets.

Built upon the Total-data-model integration (TDMI) concept, this VUQ procedure achieved three purposes: i). identify influential parameters to the quantities of interest of the boiling system through sensitivity analysis; ii). evaluate the parameter uncertainty through Bayesian inference with the support of multiple datasets; iii). quantitatively measure the agreement between solver predictions and datasets. In addition, by comparing the discrepancy between different VUQ results, the inconsistency within the closure relations can be identified. The identified inconsistency suggests the closure relations are inadequate to give accurate predictions on QoIs over the whole input space. Such inadequacy requires special attention, it is noticed that some researchers use the Gaussian process to model such inadequacy [35, 36] with support from a large amount of data.

The wall boiling closure relations studied in this work consists of widely used empirical correlations. However, the VUQ results suggest that there exists intrinsic inconsistency within the closure relation, as well as possible inconsistency with the turbulence model with wall function. Such inconsistency comes from the simplified assumption of the boiling process, as well as the negligence of interaction between different boiling phenomena. One possible approach to reduce such inconsistency is to model the boiling process with the detailed underlying physics [36-38]. Such mechanistic models, on the other hand, are highly dependent on the advanced experimental investigation of the detailed boiling process [39] and thus requires closer collaboration between model developer and experimentalist.

Through the case study, the VUQ procedure also demonstrated its potential for the multi-scale models with multiple QoIs. Be that as it may, there are still several limitations of it. First, the success of the framework is highly dependent on the availability and quality of the datasets which can cover all the QoIs of a scenario. Second, the procedure does not consider the influence of other closure relations such as turbulence modeling and interfacial forces which also could have influences on the boiling prediction. Last, the inconsistency of the closure relations is identified but not quantitatively evaluated. Given more data support, the modular Bayesian approach, which evaluate model inadequacy and model parameter uncertainty separately, could be an appropriate method to address this issue [40].

**Nomenclature**

| | | | |
|---|---|---|---|
| $A_b$ | effective bubble area fraction | *Greek symbols* | |
| $D_d$ | bubble departure diameter, m | $\alpha$ | void fraction |
| $f_d$ | bubble departure frequency, 1/s | $\theta$ | contact angle, rad |
| $\boldsymbol{g}$ | Gravity vector, m/s² | $\sigma$ | surface tension, kg/s² |
| $h$ | specific enthalpy, J/kg | $\sigma^t$ | turbulent dispersion coefficient |
| $h_{fg}$ | latent heat of evaporation, J/kg | $\lambda$ | thermal conductivity, W/(m·K) |
| $h_l$ | forced convective heat transfer coefficient, W/(m²·K) | $\mu$ | dynamic viscosity, Pa·s |
| $\boldsymbol{M}$ | interfacial force, N/m³ | $\upsilon$ | kinematic viscosity, m²/s |
| $N_a$ | nucleation site density, 1/m² | $\rho$ | density, kg/m³ |
| $Pr^t$ | turbulent Prandtl number | $\tau$ | stress tensor, kg/(m·s²) |
| $p$ | pressure, Pa | | |
| $q_{wall}$ | wall heat flux, W/m² | *Subscripts* | |
| $q_{Ev}$ | evaporation heat flux, W/m² | *sup* | superheat |
| $q_{Fc}$ | forced convective heat flux, W/m² | *l* | liquid phase |
| $q_{Qu}$ | quenching heat flux, W/m² | *v* | vapor phase |
| $T$ | temperature, K | | |
| $t$ | time, s | *Superscript* | |
| $\boldsymbol{U}$ | velocity, m/s | *t* | turbulence |
| $y^+$ | dimensionless wall distance | | |

**Acknowledgement**

This research was partially supported by the Consortium for Advanced Simulation of Light Water Reactors (http://www.casl.gov), an Energy Innovation Hub (http://www.energy.gov/hubs) for Modeling and Simulation of Nuclear Reactors under U.S. Department of Energy Contract No. DE-AC05-00OR22725. and by Nuclear Energy University Program under the grant DE-NE0008530. The authors thank Dr. Ralph Smith (NCSU, Department of Mathematics) for his valuable comments on this VUQ work.